\documentclass[prl,twocolumn,aps,nofootinbib]{revtex4}
\usepackage{amsfonts}
\usepackage{amsmath}
\usepackage[dvipsnames]{color}
\usepackage[T1]{fontenc}
\usepackage{txfonts}
\usepackage{graphicx}

\begin{document}
\title{On universal decoherence under gravity: a perspective through the Equivalence Principle}
\author{Belinda Pang}
\affiliation{Burke Institute for Theoretical Physics and the Institute for Quantum Information and Matter, M/C 350-17, California Institute of Technology, Pasadena, California 91125}
\author{Farid Ya. Khalili}
\author{Yanbei Chen}
\affiliation{Burke Institute for Theoretical Physics and the Institute for Quantum Information and Matter, M/C 350-17, California Institute of Technology, Pasadena, California 91125}

\begin{abstract}
In Nature Phys. 11, 668 (2015) (Ref.~\cite{pikovski2013universal}), a composite particle prepared in a pure initial quantum state and propagated in a uniform gravitational field is shown to undergo a decoherence process at a rate determined by the gravitational acceleration.  By assuming Einstein's Equivalence Principle to be valid, we demonstrate, first in a Lorentz frame with accelerating detectors, and then directly in the Lab frame with uniform gravity, that the dephasing between the different internal states arise not from gravity but rather from differences in their rest mass, and the mass dependence of the de Broglie wave's dispersion relation.  We provide an alternative view to the situation considered by Ref.~\cite{pikovski2013universal}, where we propose that gravity plays a kinematic role in the loss of fringe visibility by giving the detector a transverse velocity relative to the particle beam; visibility can be easily recovered by giving the screen an appropriate uniform velocity. We finally propose that dephasing due to gravity may in fact take place for certain modifications to the gravitational potential where the Equivalence Principle is violated. 
\end{abstract} 
\maketitle

\noindent {\it Introduction.--} Exciting new ideas have recently been put forward to explore the interplay between quantum mechanics and gravity using precision measurement experiments, for example testing the quantum evolution of self-gravitating objects~\cite{yang2013macroscopic}, searching for modifications to the canonical commutation relation~\cite{pikovski2012probing}, and studying propagation of quantum wavefunctions in an external gravitational field~\cite{bruschi2014spacetime,zych2012general}.  Pikovski et al.\ recently pointed out that a composite particle, prepared in an initial product quantum state between its ``center of mass'' and its internal state, will undergo a universal decoherence process with respect to its spatial degrees of freedom in a uniform gravitational field --- as exhibited by a loss of visibility in matter-wave interferometry experiments~\cite{pikovski2013universal}. More specifically, they evolved the composite particle's wavefunction in the Rindler coordinate system, and discovered a dephasing between its multiple components, at a rate proportional to the gravitational acceleration.  
%
%%adopted the Rindler coordinate system in which the composite particle's wavefunction is evolved.  From this approach, it appears that the difference in gravitational time dilation experienced by different parts of the wavefunction plays a major
%
%
%The internal structures of the particle provides a time standard for comparing the proper time difference between different interference paths. 
%

According to Einstein's Equivalence Principle (EEP), freely falling experiments cannot detect the magnitude of the absolute gravitational acceleration~\cite{will2014confrontation}.  Of course, the setup in Ref.~\cite{pikovski2013universal} is not in free fall: the screen detecting the particle is at rest in the Lab frame and accelerating in the Lorentz frame.  This means their result is not necessarily in contradiction with the EEP.  Nevertheless, it is still  interesting to explain why the dephasing, which takes place during the particle's {\it free propagation}, has a rate determined by gravity.  At this point, we note that the Rindler coordinate system is attached to the detector: calculating the dephasing of wavefunctions in that coordinate system mixes together effects due to free propagation and those due to the detector's acceleration. 

In this paper, we will first separate the propagation and detection processes. In a Lorentz frame, the internal states of the composite particle do not interact with external potentials or with each other, and are distinguished only by their rest mass. Therefore, we can treat each internal state as an independently propagating free particle species labeled by its rest mass $m$.  We will allow the screen to travel along an arbitrary transverse trajectory in one dimension with respect to the particle beam, and compute the interference pattern it registers by evaluating the particle flux over the spacetime volume spanned by the screen over measurement lifetime. The results from this perspective offer a clear alternative explanation for the loss of fringe visibility: in the Lorentz frame, the mass dependent de Broglie wave dispersion relation gives each species a mass dependent propagation velocity. This causes them to separate along the propagation axis and to arrive at the detector at different times. The detector then registers an interference pattern for each species, and if the detector is moving transversely to particle propagation (as is the case in the view of gravity being equivalent to acceleration), then the patterns from different species will be shifted along the direction of detector motion. This is equivalent to a mass dependent phase shift, and summing over all masses will result in visibility loss.

Viewed in the Lorentz frame, the final result is simple and intuitive. However, a rigorous calculation requires a quantum description of the particles that is valid in spacetime, independent of any reference frame, as well as a construction of the measurement outcome in terms of spacetime invariants. Our formalism below starts off relativistically, although we find that relativistic effects are ignorable, and that the non-relativistic limit completely reconstructs the effect found in Ref.~\cite{pikovski2013universal}. We will then go back to the Lab frame, and explicitly treat gravity as an external force field. There, we show again that the species separate due to differences in rest mass, rather than due to gravity. We point out that an explicit treatment of gravity as an external force field offer possibilities to test for EEP violations. Note that for all calculations and results we have set $\hbar=c=1$.

\noindent {\it Evolution of a Composite Particle in the Lorentz Frame.--} We will first give a frame independent quantum description of the composite particle, then examine its evolution in a Lorentz frame with Cartesian coordinates $x^\mu=(t,x,y,z)=(t,\mathbf{x})$ of a Minkowski metric. We model each species as an independent Klein Gordon field, with its field operator $\hat\phi_m(x^\mu)$ satisfying $(\square  +m^2) \hat\phi_m =0$, and expanded as
\begin{equation}\label{1} 
	\hat{\phi}_m(x^\mu)=\int \frac{\mathrm{d}^3 \mathbf{k}}{\sqrt{(2\pi)^3 2\omega_{m}(\mathbf{k})}} \; \hat{a}_m(\mathbf{k}) e^{-i\omega_m(\mathbf{k}) t +i\mathbf{k}\cdot\mathbf{x}}+h.c
\end{equation}
%where the four-vector $k^\mu$ is given by a 3+1 decomposition of 
%$%
%	k^\mu =[\omega_m(\mathbf{k}),\mathbf{k}]
%$, 
with 
$
	\omega_{m}(\mathbf{k})=\sqrt{m^2+\mathbf{k}^2}
$. 
%The mass label corresponds to different internal energy states. 
%
The commutation relations between the annihilation and creation operators are
$
	[\hat{a}_m(\mathbf{k}),\hat{a}_{m'}^\dagger(\mathbf{k'})]=\delta_{m,m'}\delta^3(\mathbf{k}-\mathbf{k'})
$.
A single particle state for mass $m$ is given by 
\begin{equation}\label{17b}
	|\Psi_m\rangle=\int \frac{\mathrm{d}^3\mathbf{k}}{\sqrt{2\omega_{m,\mathbf{k}}}}\; g_m(\mathbf{k})\hat{a}_m^\dagger(\mathbf{k})|0\rangle 
\end{equation}
where $ g_m$ transforms as a scalar between Lorentz frames, and can be viewed as the relativistic wavefunction. Note that while the interpretation of $\hat{a}_m^\dagger(\mathbf{k})|0\rangle$ as the momentum eigenstate with eigenvalue $\mathbf{k}$ is only valid for a particular Lorentz frame, $\hat{a}(\mathbf{k})$ is a well-defined operator in any frame.

When $g_m$ is limited to values of $\mathbf{k}$ such that $|\mathbf{k}| \ll m$, the quantum state is non-relativistic. Defining $f_m(\mathbf{k} ) =g_m(\mathbf{k})/\sqrt{2m}$, we construct the time-dependent wavefunction
\begin{equation}\label{psimk}
	\psi_m(t,\mathbf{x}) = \int \frac{d^3\mathbf{k}}{(2\pi)^{3/2}}f_m(\mathbf{k})e^{i\mathbf{k}\cdot\mathbf{x}}e^{-\frac{i\mathbf{k}^2t}{2m}}
\end{equation}

\begin{figure}
\includegraphics[width=0.32\textwidth]{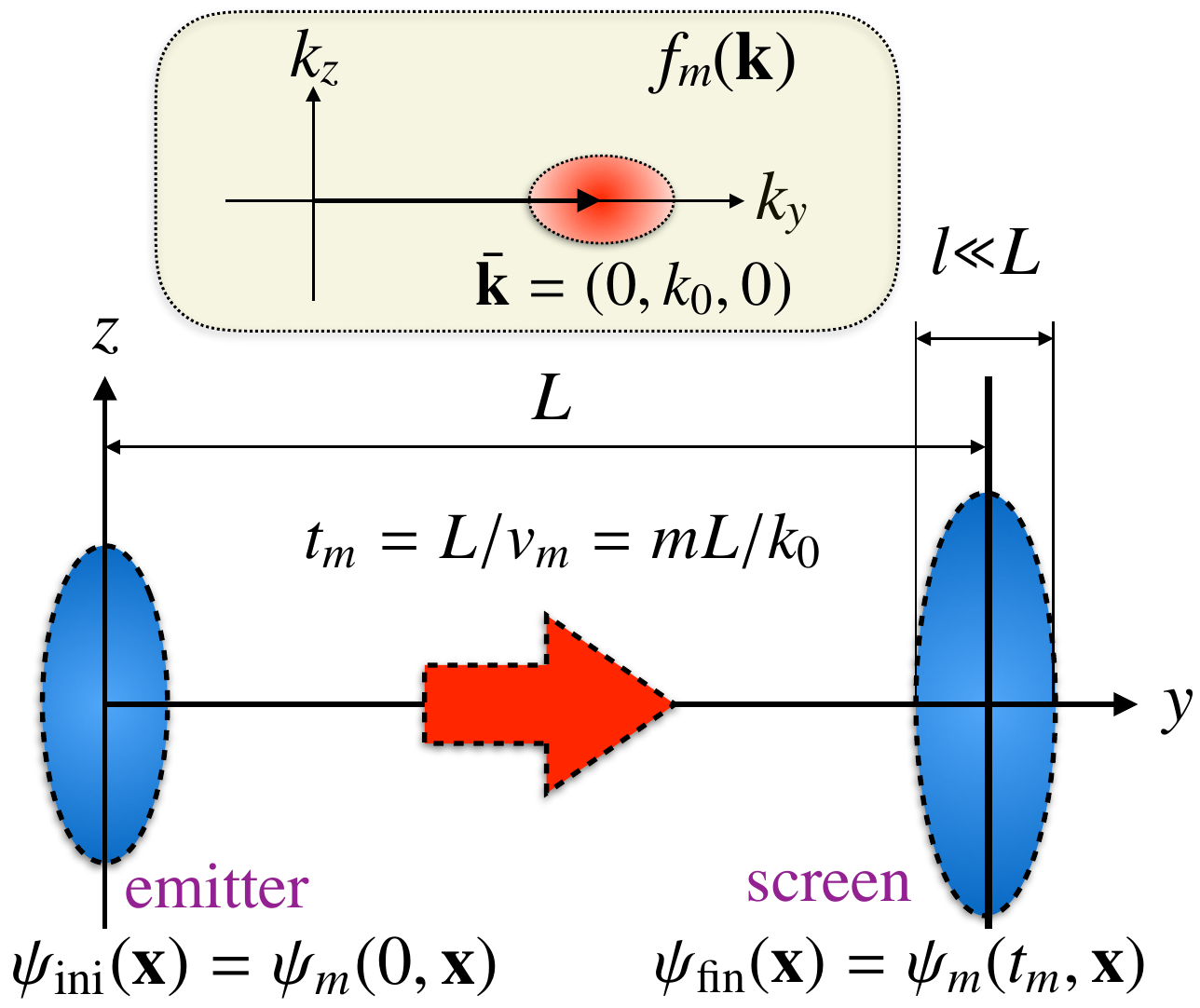}
\caption{Propagation of the wavepacket from the emitter to the screen.  The same initial wavefunction leads to the same wavefunction on the $y=L$ plane (where the screen is located), but the arrival time of the packets depend on $m$.  The inset illustrates the fact that $f_m(\mathbf{k})$ is localized around $\bar{\mathbf{k}}$.  For snapshots in time, see left panels of Fig.~\ref{fig:animals}.  \label{figprop}}
\end{figure}

At $t=0$, suppose we prepare a {\it direct-product} (i.e. a separable) state between the internal state and the translational mode of the composite particle. The translational mode corresponds to the "center of mass" degree of freedom of Ref.~\cite{pikovski2013universal}, and they contain all the information about the particle's location. This means that all species share the same initial wavefunction representing its spatial distribution, or $\psi_m(0,\mathbf{x})=\psi_{\rm ini}(\mathbf{x})$, see Fig.~\ref{figprop}, and the quantum state is prepared such that $f_m(\mathbf{k})$ in Eq.~\eqref{psimk} is $m$-independent. However, evolution in time will make $\psi_m(t,\mathbf{x})$ {\it $m$-dependent}, effectively entangling the internal and translational modes. In fact
\begin{equation}
\label{psimprop}
\psi_m(t,\mathbf{x}) = \sqrt{\frac{m}{2\pi i t}}\int d^3\mathbf{x}' e^{-\frac{m(\mathbf{x}-\mathbf{x}')^2}{2it}}
\psi_{\rm ini}(\mathbf{x}') \end{equation}
In our particular case,  suppose that $\psi_{\rm ini}(\mathbf{x})$ is spatially localized around the origin, and that the packets mainly move along the $y$ direction, which in turn means the $m$-independent $f_m(\mathbf{k})$ is localized near $\bar{\mathbf{k}} = (0,k_0,0)$ (inset of Fig.~\ref{figprop}).  The mean velocity of the $m$ species will be $v_m = k_0/m$, and the time it takes the center of the $m$ packet to reach $y=L$ is 
$
t_m \equiv L/v_m = m L/k_0
$.
Inserting $t_m$ into Eq.~\eqref{psimprop}, we obtain
\begin{equation}
\label{diff}
\psi_m(t_m,\mathbf{x}) =\sqrt{\frac{k_0}{2\pi i L}}\int d^3\mathbf{x}' e^{-\frac{k_0(\mathbf{x}-\mathbf{x}')^2}{2iL}}
\psi_{\rm ini}(\mathbf{x}')   \equiv \psi_{\rm fin}(\mathbf{x})\,.
\end{equation}
In other words, even though $v_m$ and $t_m$ are both $m$-dependent, wavefunctions upon arrival at the screen are not. In fact, Eq.~\eqref{diff} is simply the Fresnel diffraction formula: the diffraction pattern on the screen depends on the initial wavefunction, the propagation distance, and the de Brogelie wavelength, none of which are $m$ dependent.  We assume the packet's size to be much less than $L$ throughout the propagation (see Fig.~\ref{figprop}).
\noindent {\it Detector trajectory in a Lorentz Frame.--}   Having prepared a freely propagating beam along $y$, we now characterize the screen which is moving arbitrarily along $z$. We parametrize the {\it central pixel of the screen} 
\begin{equation}
	x_{\rm cs}^{\mu} =  [t_{\rm cs}(\tau),0,L,z_{\rm cs}(\tau)]
\end{equation}
by its proper time $\tau$.   We denote the instantaneous 3-velocity and Lorentz factor by
$	\beta(\tau)= dz_{\rm cs}/dt_{\rm cs}$ and $\gamma(\tau )= dt_{\rm cs}/d\tau$ respectively. The proper acceleration is given by 
$	g= \gamma^2 \dot\beta$.
%The proper reference frame is established by sending spatial geodesics away from the central pixel, along the vectors $e_0^\mu = u_{\rm cs}^\mu$, and 
%\begin{equation}
%	e_1^\mu = (0,1,0,0),\; 
%	e_2^\mu = (0,0,1,0),\;
%	e_3^\mu = (\gamma\beta,0,0,f\gamma),
%\end{equation}
%which form an orthonormal tetrad.
A {\it proper reference frame}~\cite{misner1973gravitation} can be established for the central pixel of the screen has coordinates $ (\tau, X,Y,Z) $, which maps to Minkowski as 
\begin{equation}\label{coordtrans}
	x^\mu(\tau,X,Y,Z) = \left[t_{\rm cs}(\tau)+\beta\gamma Z,X,Y+L,z_{\rm cs}(\tau)+ \gamma Z\right]\,
\end{equation}
Note that this only covers a region around the worldline of the central pixel. The metric can be written as
\begin{equation}
	ds^2 =  -(1+g Z)^2d\tau^2 +dX^2 + dY^2+ dZ^2
\end{equation} 
%Note that in this coordinate system, 
In this coordinate system, pixels on the screen are parametrized spatially by by $(X,Z)$, with $ Y=0 $ and proper time $(1+gZ)\tau$. 

\noindent {\it The Measurement Process.--} 
For each species, we measure the number of particles captured by the screen per area over the lifetime of the experiment. We denote this quantity by $\sigma_m$, and our final outcome $\sigma$ is the sum over all species. Defining $ \hat{\phi}_{m,+} $ and $ \hat{\phi}_{m,-} $ as the positive and negative frequency components of $ \hat{\phi}_m $, we first introduce the 4-current operator
\begin{align}
	\hat{j}_m^\nu(x^\mu)
	=i\left[\partial^\nu\hat{\phi}_{m,-}(x^\mu)\right]\hat{\phi}_{m,+}(x^\mu)+h.c.
\end{align}
For a particular quantum state $|\Psi\rangle$,  $ \langle\Psi|\hat{j}_m^\nu(x^\mu)|\Psi\rangle$ represent the 4-probability current for species $m$. When the quantum state is non-relativistic, the probability density and flux become
\begin{equation}
\langle\Psi_m|\hat{j}_m^\mu(t,\mathbf{x})|\Psi_m\rangle =\left[|\psi_m|^2,
\frac{1}{2mi}\left(\psi_m^* \nabla \psi_m -\psi_m \nabla\psi_m^*   \right)
\right]
\end{equation}
 For the spacetime volume $V$ spanned by the proper area of a patch on the screen and the measurement duration, the number of particles captured by the screen is given by the flux integral
\begin{align}\label{5}
	N_m[V]=
	\int_{V} \;
	d\Sigma_\nu\langle\Psi_m|\hat{j}_m^\nu|\Psi_m\rangle\,.
\end{align}

To be precise, this relies on the assumption that each space-time pixel on the screen operates independently. Because $d\Sigma_\nu =\left[0,0,(1+gZ) dXdZd\tau,0 \right]$, we can write $N_m[V] =\int_{V} dXdZd\tau\;\dot \sigma_m(\tau,X,Z)$, where
\begin{equation}\label{sigmadotdef}
\dot{\sigma}_m (\tau,X,Z)  = {(1+gZ)\langle\Psi_m|\hat{j}_m^y(x^\mu)|\Psi_m\rangle}
\end{equation}
Eq.~\eqref{sigmadotdef} has the interpretation of number of particles per proper area per coordinate time $\tau$. Integrating over measurement time, we have
\begin{equation}
\label{sigmamint}
\sigma_m(X,Z)=\int d\tau\;\dot{\sigma}_m(\tau,X,Z)
\end{equation}
Finally, the total count per area for mass distribution $P_m$ is given by 
\begin{equation}\label{sigma}
\sigma(X,Z)=\int dm\;P_m\sigma_m(X,Z)
\end{equation}
%where
%\begin{equation}
%\sigma_m(X,Z)=\int d\tau\;\dot{\sigma}_m(\tau,X,Z)
%\end{equation}
%We note that the integration is over the coordinate time and not each pixel's proper time, except for those with $ Z=0 $. There is a subtlety in the physical interpretation of our measurement scheme whether we use the coordinate time or the proper time. In the first case, an observer at $ Z=0 $ turns each pixel on and off using clocks synchronized to his own proper time. In the second case, the pixels are turned on and off by some internal clock. However, because we will assume that our measurement duration is long enough such that the probability flux is zero at the beginning and end times, we can extend our limits of integration in either case to $ \pm\infty $, and this subtlety will not affect our results.

\noindent {\it The Interference Pattern and Loss of Visibility.--} 
We will now make use of our simplifying assumptions and calculate $ \dot{\sigma}_m $. First, because $f_m(\mathbf{k})$ is localized around $\bar{\mathbf{k}}$, we have $\partial_y \psi_m \approx ik_0\psi_m$ at any given time [see Eq.~\eqref{psimk}], and 
%
% First, we assume that momentum is mostly in the $ y $ direction and centered about $ k_0 $, so that $ \kappa\ll k_0 $, where $ \kappa=|\mathbf{k}-k_0\hat{e}_y| $ is the size of the wavepacket in $ k $-space. Then with error $\mathcal{O}(\kappa/k_0) $
\begin{align}
\label{sigmadot}
\dot{\sigma}_m(\tau,X,Z)
=&v_m(1+gZ)
|\psi_m^2\left(x^\mu(\tau,X,0,Z)\right)|
\end{align}
%and
%\begin{equation}
%\label{sigmam}
%\sigma_m(X,Z) = \int v_m(1+gZ)
%|\psi_m^2\left(x^\mu(\tau,X,0,Z)\right)| 
%d\tau
%\end{equation} 
%where $ \psi_m(t,\mathbf{x}) $ is given by Eq.~(\ref{24}).
%Next, from Eq.~\eqref{coordtrans}, we solve 
%\begin{equation}
%z(t,Z)  = \tilde{z}_{\rm cs}(t) + {Z}/{\gamma} +O({gZ^2})\,,\quad 
%{dt}/{d\tau} = \gamma (1+gZ)
%\end{equation}
Using Eq.~\eqref{coordtrans}, we can express the Minkowski coordinates in $|\psi^2_m(x^\mu)|$ explicitly in terms of $(X,Z)$. We also change the integration variable in Eq.~\eqref{sigmamint} from $\tau$ to $t$, which we can now write as
%where we've defined
%\begin{equation}
%\tilde{z}_{\rm cs}(t)=z_{\rm cs}[\tau_{\rm cs}(t)],\quad \tau_{\rm cs}[t_{\rm cs}(\tau)]=\tau
%\end{equation}
%In words, here $ \tau_{\rm cs}(t) $ is the inverse of $ t_{\rm cs}(\tau) $, and $ \tilde{z}_{\rm cs} $ is an %explicit function of $ t $. We convert Eq.~\eqref{sigmamint} into an integral in $t$, instead of $\tau$, %writing
\begin{equation}\label{sigmam}
\sigma_{m}(X,Z) =v_m \int \gamma\left|\psi_m^2\left[t,X,0,\tilde z_{\rm cs}(t)+ \gamma^{-1} Z\right] \right|^2dt
\end{equation}
In the above, we've defined $\tilde{z}_{\rm cs}(t)=z_{\rm cs}[\tau_{\rm cs}(t)]$, where $\tau_{\rm cs}(t)$ is the inverse of $ t_{\rm cs}(\tau) $ and is the proper time of the central pixel as a function of the Minkowski coordinate time. Then $ \tilde{z}_{\rm cs} $ is an explicit function of $t$ that gives the central pixel's $z$ position in Minkowski time.

\begin{figure}
			\includegraphics[width=0.475\textwidth]{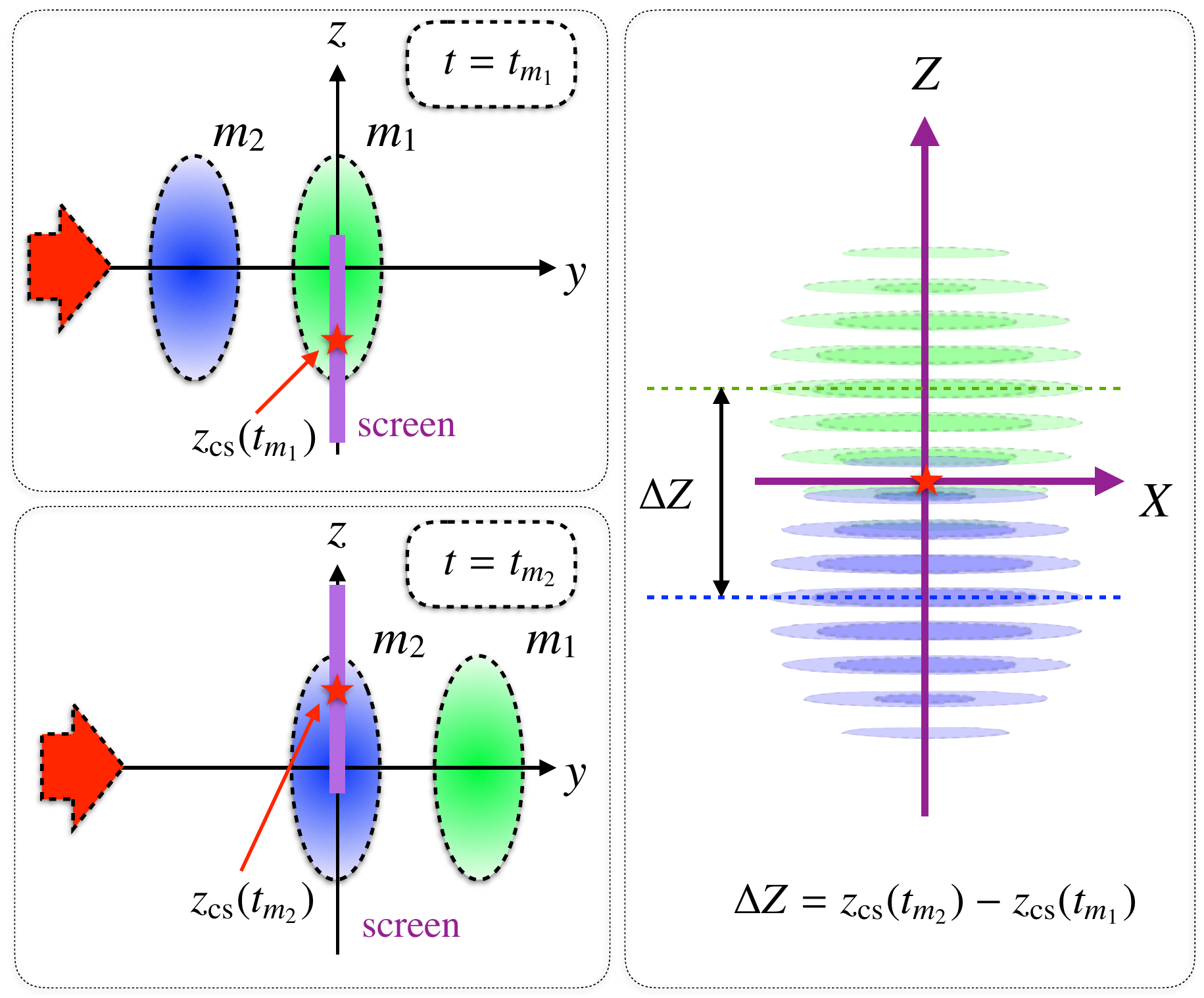}
	\caption{Snapshots taken upon arrival of $m_1$ and $m_2$ ($m_1<m_2$) packets at the screen, at $t_{m_1}$ (upper left panel) and $t_{m_2}$ (lower left panel), respectively (separation between the packets highly exaggerated). Positions of the screen differ at these moments (with central pixel labeled by red star), causing a shift in the interference pattern registered by the screen, which is best viewed in the central pixel's proper reference frame (right panel).  \label{fig:animals}}
\end{figure}

We will now map $\psi_m$ in Eq.~\eqref{sigmam} to $\psi_{\rm fin}$, which is mass independent.  Because the packet is localized within $ l \ll L $ along $ y $, the particle will be detectable by the screen over a finite measurement duration, 
$t_m-{l}/{v_m} <  t < t_m+{l}/{v_m}$, where recall that $t_m$ is the mass dependent arrival time at the screen for species $m$. 
During this time, the packet remains rigidly moving along $y$ with $v_m$~\footnote{Note that $t_m$ and $v_m$ each also has a range due to the mass uncertainty, although those  uncertainties cause only a small increase in the overall range.}.  
Since the packet does not evolve in shape as it crosses the screen, and suppressing $x$ and $z$, we can equate $\psi_m(t_m+\epsilon,y)$ -- where $t$ varies, with $\psi_m(t_m,y-v_m\epsilon)$ -- where $y$ varies, for $|\epsilon| \stackrel{<}{_\sim}{l}/{v_m}$. But $\psi_m(t_m)=\psi_{\rm fin}$, so we can rewrite Eq.~\eqref{sigmam} as
%\begin{equation}\label{34}
%|\psi^2_m(t_m+\epsilon,x,y,z)|=|\psi^2_{\rm{fin}}(x,y-v_m\epsilon,z)|\,,\; |\epsilon| \stackrel{<}{_\sim}{l}/{v_m}\,. 
%\end{equation}
%Here $\tilde\psi_{\rm fin}$ can still differ significantly from $\tilde\psi_{\rm ini}$, if $L > k_0/\kappa^2$.  
%With the above, we can rewrite Eq. (\ref{sigmam}) as
\begin{align}
\sigma_m(X,Z)
=\int dw\; \gamma\left| {\psi}^2_{\rm fin}(X,L-w,\tilde z_{\rm cs}(t_m+{w}/{v_m})+\gamma^{-1} Z)\right|\label{sigmamw}
\end{align}
Simply put, since each packet arrives on the screen at different moment of time, while the screen is moving, the pattern registered by the screen will be $m$ dependent, as shown in Fig.~\ref{fig:animals}.  More specifically, the screen pixel $(X,Z)$ samples the wave packet along a trajectory, parametrized by 
\begin{equation}\label{sampling}
[y,z]= [L-w, \tilde z_{\rm cs}(t_m+w/v_m) +\gamma^{-1} Z]
\end{equation} 
Note that $t_m$ and $v_m$ both depend on mass, which will lead to a mass-dependent interference pattern, through $\tilde z_{\rm cs}$.  Since the range of $w$ is between $(-l,l)$, then for $ l\ll L $, the difference in $w/v_m$ caused by mass difference will be much less than that in $t_m$, so we can ignore $w/v_m$ in $z$. 

So far, the {\it screen} is still relativistic (although the quantum state is not). To clearly demonstrate the loss of visibility, let us assume low velocities for the screen, $ \gamma \approx 1 $. Additionally, we assume separability such that $\psi_{\rm fin} (x,y,z) =\psi_{\rm fin}^x (x) \psi_{\rm fin}^y (y) \psi_{\rm fin}^z(z)$
%\begin{equation} 
%\psi_{\rm fin} (x,y,z) =\psi_{\rm fin}^x (x) \psi_{\rm fin}^y (y) \psi_{\rm fin}^z(z)
%\end{equation}
In this case,
\begin{equation}\label{zposition}
\sigma_m(X,Z) =  |{\psi}_{\rm fin}^x(X)|^2 \left[\int dw |{\psi}^y_{\rm fin} (w)|^2\right] \left|{\psi}^z_{\rm fin}[\tilde  z_{\rm cs}(t_m) + Z]\right|^2
\end{equation}
Suppose $\psi_{\rm fin}^z$ contains an interference pattern with visibility $V$, so that locally around $z\approx \tilde{z}_{\rm cs} (t_m)$ we have
\begin{equation}\label{psisq}
|\psi_{\rm fin}^z(z) |^2 = C \left[1+V\cos(\alpha z+\phi)\right]
\end{equation}
where $\alpha$ is the wavenumber of the spatial oscillation. We can ignore $\phi$ without loss of generality, and write
\begin{equation}
\label{sigmazp}
\sigma(Z) \propto \int P_m \left[1+V \cos[\alpha z_{\rm cs}(t_m)+Z]\right] dm
\end{equation}
We can expand $\tilde{z}_{\rm cs}(t_m)$  around $t_{\bar{m}}$, which is the time of arrival of the particle having average mass $\bar{m}=\int m P_m dm$.
%We can expand
%\begin{align}
%\tilde{z}_{\rm cs} (t_m)= \tilde{z}_{\bar m} +  \dot{\tilde{z}}_{\bar m} (t_m - t_{\bar m})+ { \ddot {\tilde{z}}_{\bar m}
%	(t_m - t_{\bar m})^2}/{2}
%\end{align}
%with
%${\tilde{z}}_{\bar m} \equiv \tilde{z}_{\rm cs} (t_{\bar m})$,
%${\dot {\tilde{z}}}_{\bar m} \equiv \dot {\tilde{z}}_{\rm cs} (t_{\bar m})$, 
%${\ddot {\tilde{z}}}_{\bar m} \equiv \ddot {\tilde{z}}_{\rm cs} (t_{\bar m})$,
%and 
%\begin{equation}
%\bar m  \equiv \int m P_m dm,\quad \Delta m^2 \equiv {\int (m-\bar m)^2 P_m dm}
%\end{equation} 
%This leads to 
%\begin{align}
%&\cos[\alpha z_{\rm cs}(t_m)+Z] \nonumber\\
%=&
%\left[1-\frac{\alpha^2 {\dot z}_{\bar m}^2 (t_m-t_{\bar m})^2}{2}\right] 
%\cos [\alpha z_{\bar m}+Z] \nonumber\\
%- &
%\alpha \left[ {\dot z}_{\bar m}(t_m - t_{\bar m})+ \frac{{\ddot z}_{\bar m}
%	(t_m - t_{\bar m})^2}{2}\right]\sin [\alpha z_{\bar m}+Z]
%\end{align}
Using the subscript $\bar{m}$ to denote quantities of the average mass particle, Eq.~\eqref{sigmazp} becomes
\begin{align}
\sigma(Z) \propto 1+ V' \cos[\alpha \tilde{z}_{\bar{m}}+ Z +\phi']
\end{align}
where, defining $\Delta m^2={\int (m-\bar m)^2 P_m dm}\,$ as the mass variance and using a dot (as in $\dot{z}$) to signify the time derivative, we have
\begin{align}
\frac{V'}{V} = 1-\frac{\alpha^2 {\dot {\tilde{z}}}_{\bar{m}}^2 t_{\bar m}^2  }{2 }
\left(\frac{\Delta m}{\bar m}\right)^2
\,,\quad \phi' = \frac{\alpha {\ddot {\tilde{z}}}_{\bar m} t_{\bar m}^2 }{2}\left(\frac{\Delta m}{\bar m}\right)^2
\label{visphase}
\end{align}  
For large $t_{\bar{m}}$, we cannot make the approximations that led to Eq. (\ref{visphase}). Instead, from Eq. (\ref{sigmazp}), we have
\begin{equation}\label{sigmazrev}
\sigma(Z)\propto 1+\Re\left\{e^{i\alpha z_{\bar{m}}}\int P(m)e^{i\delta\phi_m}\;dm\right\}
\end{equation}
where
\begin{equation}
\delta \phi_m=\alpha[z_{\bar{m}}-z(t_m)]\approx \alpha \dot{z}_{\bar{m}} \frac{(\bar{m}-m)L}{ k_0}
\end{equation}
The second term in Eq.~\eqref{sigmazrev} is the real projection of a weighted sum of phasors. The visibility is simply the magnitude of the overall phasor, given by
\begin{equation}\label{totalvis}
V'=\left|\langle e^{i\delta\phi_m}\rangle\right|
\end{equation}
where the brackets signify an average over the mass distribution. For a finite number of mass components, one can always find a time $t_{\bar{m}}$ for which the difference between any pair of $\delta\phi_m$ is an integer multiple of $2\pi$. At this value, $V'=1$ and visibility is recovered, corresponding to the "revival" of Ref.~\cite{pikovski2013universal}.

\noindent {\it A Double Slit Experiment in a Uniformly Accelerating Lab.--} We now consider the specific case of an initial superposition of the translational mode, or a double slit experiment, as in Ref.~\cite{pikovski2013universal}. The initial wavefunction is given by 
\begin{equation}
	\psi_{\mathrm{ini}}(x,y,z)\propto \psi_{\rm ini}^x(x)\psi_{\rm ini}^y(y)[\delta(z-z_1)+\delta(z-z_2)]
\end{equation}
this leads to $\psi_{\rm fin}(z)$ with a perfect contrast, with 
\begin{equation}
V=1\,,\quad \alpha = k_0 (z_1-z_2)/L
\end{equation}
We also have 
%Then by Eq.'s (\ref{21'}) and (\ref{25}), Eq (\ref{psisq}) takes the form $ |\psi_z(z)|^2=1+\cos\left(\frac{k_0\Delta z}{L}z\right) $, where $ \Delta z=z_1-z_2 $, and we can identify $ \alpha=\frac{k_0\Delta z}{L} $, $ C=V=1 $. Then Eq. (\ref{zstz}) becomes
$
 	\tilde{z}_{\rm cs}(t)=gt^2/2$,
$\dot {\tilde{z}}_{\bar m} = gt_{\bar m}$ and $\ddot {\tilde{z}}_{\bar m} = g$, which, for short time scales, leads to an integrated visibility of
\begin{equation}\label{visds}
V'=1-{\left[g(z_1-z_2) \Delta m \right]^2}t_{\bar{m}}^2/2,
\end{equation}
For a thermal distribution of $ N $ harmonic degrees of freedom at high temperature $ T $, $ \Delta m=k_B T \sqrt{N} $, and Eq. (\ref{visds}) becomes
\begin{equation}
V'(t_{\bar{m}})\approx e^{-(t_{\bar{m}}/\tau_{\mathrm{dec}})^2},\;\quad \tau_{\mathrm{dec}}=\sqrt{\frac{2}{N}}\frac{1}{k_BT g|z_2-z_1|}
\end{equation}
which exactly reproduces Eq.~(3) of Ref.~\cite{pikovski2013universal}. For longer times, we consider the full expression in Eq. (\ref{totalvis}), where now
\begin{equation}
\delta\phi_m=gt_{\bar{m}}(\bar{m}-m)(z_2-z_1)
\end{equation}
In the language of Ref.~\cite{pikovski2013universal}, $(\bar{m}-m)=(\langle H_0\rangle -H_0)$ and $g(z_2-z_1)t_{\bar{m}}=\Delta \tau $, where $\Delta\tau$ is the difference in proper times between superposed paths of the particle. Then, we obtain
\begin{equation}\label{fullvisloss}
V'=\left|\left\langle e^{iH_0 \Delta \tau}\right\rangle\right|
\end{equation}
which reproduces the result in Eq (4) of Ref.~\cite{pikovski2013universal} for the double slit system.

\noindent{\it Origin of the loss in visibility.--} While our predicted loss of visibility in Eq's (\ref{visphase}) and (\ref{fullvisloss}) are the same as that of Ref.~\cite{pikovski2013universal}, we interpret this effect as being unrelated to gravity. The source of dephasing is the mass dependent propagation velocities, which causes the different species to separate along $ y $. This separation implies that a particular pixel on the moving screen will sample the wavefunction $\psi_{\rm fin}(\mathbf{x})$ along mass dependent trajectories as given by Eq.~\eqref{sampling}, effectively evaluating $\psi_{\rm fin}(\mathbf{x})$ at mass dependent positions along $z$ as shown in Eq.~\eqref{zposition} and in the left panels of Fig.~\ref{fig:animals}. In the screen's own reference frame, this means that different species land on different locations, the causing mass-dependent shifts of their interference patterns along $ Z $, by $\tilde{z}_{\mathrm{cs}}(t_{m})$ [Fig.~\ref{fig:animals}, right panel], which in turn smears out the pattern. In this way, the loss of visibility is directly related to the transverse {\it velocity} of the screen~[Eq.~\eqref{visphase}], instead of {\it acceleration}.  In the situation considered by Ref.~\cite{pikovski2013universal}, gravity happens to supply such a transverse velocity, thereby making the decrease in visibility dependent upon the gravitational acceleration.  However, if we give the screen a uniform velocity in the Lab frame (with gravity) that matches the velocity at which the packets fall in the lab frame, there will be no loss of visibility. {\it Vice versa}, even in absence of gravity, any motion of the screen transverse to the beam's propagation direction as the packets land will lead to a loss of visibility. Ultimately, {\it it is the mass dependence of the de Broglie wave's dispersion that led to wavepacket separation among different species, and subsequently the transverse motion of the screen that led to smearing.}  

Thus, our formalism offers an alternative perspective that the loss of visibility is a is a {\it kinematic effect} instead of a gravitational one. This becomes clear in our calculations is because we consider the full interference pattern imprinted by each species onto the screen, from which we notice that patterns are shifted along the direction of detector motion. In contrast, Ref.~\cite{pikovski2013universal} considered two point quantum correlations, which makes it difficult identify this shift.
%
%\begin{equation}
%\Delta Z(m) = - \frac{g t_m^2}{2} =-\frac{gL^2}{2v_m^2} 
%\end{equation}
%and we have
%\begin{equation}
%z_{\bar{m}}=Z+\frac{1}{2}gt_{\bar{m}}^2,\;\quad \dot{z}_{\bar{m}}=gt_{\bar{m}}
%\end{equation}
%so that
%\begin{equation}\label{visds}
%V'=1-\frac{g\Delta z t_{\bar{m}}}{2}\Delta m^2,\;\quad \delta\phi=\frac{g\Delta z}{2}\Delta m^2
%\end{equation}

\begin{figure}
			\includegraphics[width=0.315\textwidth]{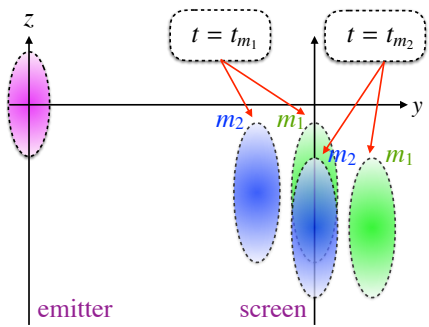}
	 \caption{Snapshots of wave packets in the the Lab frame with uniform gravity, upon arrival of the $m_1$ and $m_2$ packets on the screen. Note that the separation of packets are along the initial velocity, and gravity does not cause separation.  The packets land on different parts of the screen, causing the same smearing of the interference pattern as shown on the right panel of Fig.~\ref{fig:animals}. If the screen were to drop at the same velocity as the packets as they land, the visibility will be recovered. \label{fig:heis}}
\end{figure}

\noindent{\it Direct calculation in gravitational field.--} That mass is the source of dephasing instead of gravity can also be shown when the gravitational field is treated explicitly. Calculation in Lorentz frame showed that the thought experiment can be treated non-relativistically as long as the internal energy is accounted for in the rest mass.  Quantizing the classical Hamiltonian for a particle in free fall, and extending this to account to different mass eigenstates, we write following Hamiltonian for multiple particle species:
\begin{equation}\label{gravH}
\hat{H} = {\hat{\mathbf{p}}^2}/({2\hat M}) + \hat M \mathbf{g}\cdot \hat {\mathbf{x}}
\end{equation}
More specifically, we have a Hilbert space $\mathcal{H} =\oplus_m \mathcal{H}_m$, where each $\mathcal{H}_m$ is the eigensubspace of the mass operator $\hat M$ with eigenvalue $m$. The operators $\hat{\mathbf{x}}$ and $\hat{\mathbf{p}}$ are the position and momentum operators in each $\mathcal{H}_m$, with  $[\hat M,\hat{\mathbf{x}}]=[\hat M,\hat{\mathbf{p}}]=[\hat M,\hat H]=0$, and $\mathbf{g}$ is the gravitational acceleration.  Heisenberg equations are given by  $\dot{\hat{\mathbf{x}}} =\hat{\mathbf{p}}/\hat M$ and 
$\dot{\hat{\mathbf{p}}} =-\hat M\mathbf{g}$, and 
\begin{align}
\label{eqx}
\hat{\mathbf{x}}(t) &= \hat{\mathbf{x}}_0 + \hat{\mathbf{p}}_{0} t/{\hat M} - \mathbf{g} t^2/2\,. 
\end{align}
In this picture, it is very clear that the separation of the packets is due to the spectrum of $\hat M$, while the effect of gravity is common for all species.  In our particular thought experiment, $\hat p_y(0)$ is given an initial distribution peaked around $ k_0$, there will be a $y$-direction separation of the packets, which makes the $z$-patterns shift as they arrive on a screen with fixed $y=L$ (see Fig.~\ref{fig:heis}). 

The loss of visibility discussed in this paper relies on the equivalence between gravity and acceleration -- hence it relies on EEP. The explicit gravitational field treatment above is useful as a starting point for testing violations of EEP, which may have implications in the quantum regime that cannot be tested by classical systems, as discussed in ~\cite{viola1997testing} and ~\cite{zych2015qEEP}.  We note that while Eq. (\ref{gravH}) does not assume EEP {\it a priori}, it is implicit in the fact that the same operator $ \hat{M} $ appears as both inertial and gravitational mass. A test of EEP will require modifications to the Hamiltonian in Eq.~\eqref{gravH}.

\noindent {\it Conclusions.--} Having treated the matter-wave interferometry thought experiment both in Lorentz Frame with accelerating detectors and in the Lab frame with gravity, in both cases we offer the point of view that dephasing between different internal states do not arise from gravity, but instead from the mass dependence of the deBroglie wave's dispersion and the relative transverse motion of the detector. Furthermore, the dephasing we calculate from this perspective is the same as that predicted by Ref.~\cite{pikovski2013universal} using their perspective of time dilation. We comment that the authors of Ref.~\cite{pikovski2013universal} hold the position that the effect they describe is due to time dilation, and thus a different effect than discussed here. We note that in the Lorentz frame, we have assumed EEP to be valid, and in the Lab frame, EEP is already embedded in the Hamiltonian. In order to study the implications of EEP in quantum systems and to test for its violations in this regime, we will have to consider a more general Hamiltonian in the Lab frame
\begin{equation}
\hat H = {\hat{\mathbf{p}}^2}/{(2\hat M)} - \hat{\mathbf{G}}\cdot\hat{\mathbf{x}}
\end{equation}
where $\hat{\mathbf{G}}$, the gravitational force, is no longer given by $\hat M \mathbf{g}$.  In this case, the packets of multiple mass components will separate due to both the spectrums of $ \hat{M} $ and $ \hat{G} $, and now gravity will cause packet separation. It is also plausible that preparation of novel quantum states can reveal more structures in the operator $\hat G$ that could otherwise be revealed by a classical experiment~\cite{viola1997testing}.

\noindent{\it Acknowledgements.--} Research of Y.C.\ and B.P.\ are supported by the Institute for Quantum Information and Matter, as we as NSF Grants PHY-1404569 and PHY-1506453. Research of F.K. was supported by  LIGO NSF Grant PHY-1305863 and Russian Foundation for Basic Research Grant 14-02-00399. We thank C.\ Brukner, I.\ Pikovski, Y. \ Ma, B.L.\ Hu and O.\ Romero-Isart for exciting discussions. 

\bibliographystyle{apsrev}
\bibliography{references}
\end{document}